\documentclass[12pt]{iopart}
\usepackage{graphicx}
\usepackage{colordvi}
\usepackage[small, bf]{caption}

\newcommand{\ve}[1]{\mbox{\boldmath $#1$}}

\begin{document}
\title[Robustness Leads Close to the Edge of Chaos in Coupled Map Networks]{Robustness Leads Close to the Edge of Chaos in Coupled Map Networks:  toward the understanding of biological networks}

\author{Nen Saito}
\address{
Graduate School of Science, Osaka University, Toyonaka, Osaka 560-0043,
Japan}
\address{
Cybermedia Center, Osaka University, Toyonaka, Osaka 560-0043, Japan}
\address{
Graduate School of Arts and Sciences
The University of Tokyo 3-8-1 Komaba, Meguro-ku Tokyo 153-8902, Japan
\footnote{(present address)}}
\ead{saito@complex.c.u-tokyo.ac.jp}
\author{Macoto Kikuchi}
\address{Cybermedia Center, Osaka University,
Toyonaka, Osaka 560-0043, Japan}
\address{Graduate School of Science, Osaka University, Toyonaka, Osaka 560-0043,
Japan
}
\address{Graduate School of Frontier Biosciences, Osaka University, Suita, Osaka 565-0871, Japan}

\ead{kikuchi@cmc.osaka-u.ac.jp} 
\date{\today}
\begin{abstract}
Dynamics in biological networks are in general robust against several perturbations. We investigate a coupled map network as a model motivated by gene regulatory networks and design systems which are robust against phenotypic perturbations (perturbations in dynamics), as well as systems which are robust against mutation (perturbations in network structure).  To achieve such a design, we apply a multicanonical Monte Carlo method. 
Analysis based on the maximum Lyapunov exponent and parameter sensitivity shows that systems with marginal stability, which are regarded as systems at {\it the edge of chaos}, emerge when robustness against network perturbations is required. This emergence of {\it the edge of chaos} is a self-organization phenomenon and does not need a fine tuning of parameters.
\end{abstract}
\maketitle

\section{Introduction}

Complex dynamical behaviors on a network can be found in a variety of biological networks, such as gene regulatory networks, neural networks and food-web.
Such systems share a common characteristic: observed dynamics are robust against disturbance introduced in its dynamics, as well as against disturbance in its network~\cite{visser2003perspective,li2004yeast,wagner2005robustness}.
For example, gene expression patterns obtained thorough transcription-translation regulations are kept stable in spite of extrinsic noises (i.e., perturbations in dynamics) and mutations (i.e., perturbations in a network).
It seems reasonable to think that robustness against environmental perturbations has been evolutionarily developed for adapting to noisy environments. There are also several advantages in having mutational robustness - it buffers against deleterious mutations.
Recently, robustness in biological networks has attracted much attention of many researchers and has been thought to be one of fundamental properties of life~\cite{visser2003perspective,li2004yeast,wagner2005robustness,masel2010robustness}.

This point of view naturally gives rise to the question:
what kind of system emerges when only robustness is required ?
The answer to this question will be helpful for understanding the design principle of living systems.
In this paper, we investigate a coupled chaotic map network motivated by gene regulatory networks and show that systems at the {\it edge of chaos} are selected with only the requirement of robustness against network perturbations.

It has long been hypothesized that living
systems favor the {\it edge of chaos}, where stability and
chaoticity coexist.
Originally, Kauffman~\cite{kauffman1993origins} introduced the Boolean
network model (N-K model) as a model of a gene regulatory
network, and proposed the hypothesis
that 
living systems prefer 
 {\it the edge of chaos} because it allows systems to have complex behaviors~\cite{kauffman1993origins}.
Here we propose an alternative scenario, specifically that the requirement of having robustness against network perturbations drives living systems to the edge of chaos, regardless of whether or not staying at {\it the edge of chaos} is beneficial for living systems. In other wards, the edge of chaos can emerge as a byproduct of the robustness.

\section{Model}
We propose a coupled map system motivated by gene regulatory networks.
Unlike the N-K model, each element in this model has its own
dynamics.
Assuming that $x^t _i \in \{ -1\le x_{i }^{t}\le 1 \}$ is the gene expression of the {\it i}-th gene at time step
$t$, the single gene dynamics are written as
$x_i^{t+1}=G(x_i^t)$.
These dynamics mimic multiple processes in an expression of a single
gene.
In the presence of $N$ genes, the dynamics of $x_i^{t}$ are expressed as
\begin{equation}
  x_i^{t+1}=(1-\epsilon)G(x_i^t)+\epsilon \sum_j ^N W_{ij} G(x_j^t),
\label{eq:model}
\end{equation}
where $\epsilon$ is a coupling constant. 
$W_{ij}$ describes the strength of the interaction acting from gene 
{\it j} on gene {\it i};
and $W_{ij} \in [0,1]$ satisfies both conditions $W_{ii}=0$ and $\sum _j ^N
W_{ij}=1$ for each $i$.
We call the matrix $W$, whose $ij$ element is $W_{ij}$, a network.
Here we choose the logistic map $g(x,a)=1-ax^2$ as $G(\cdot)$.
We use the model parameters $a$ and $\epsilon$ as $(a,\epsilon)=(1.8,0.1)$. This choice indicates that a single disconnected gene exhibits chaotic dynamics.
A reason of this choice of $G(\cdot)$ is that a single gene expression is expected to be complex due to multiple processes underlying it
We impose an additional constraint on $W$: the number of input links
$k$ to each gene is fixed.
We note that in the case of $W_{ij}=1/(N-1)$ for all $(i,j)$, the system becomes the globally coupled map (GCM)~\cite{kaneko1990clustering} and it shows
highly chaotic behaviors at the parameters $(a,\epsilon)=(1.8,0.1)$.
In contrast to the N-K model, variables of this model take continuous
values and a linear
stability analysis can be applied.

In this model, the network $W$ is regarded as a genotype
while the attractor of dynamics $\ve{x}^t$ is regarded as a phenotype.
Our goal is to design networks under the two different design principles:
robustness against phenotypic perturbations
(i.e., perturbations in the dynamics of gene $\ve{x}$) and
robustness against genotypic perturbations (i.e., perturbations on
network $W$). In both cases, only network $W$ is tuned.

\section{Design of Robust Network Against Perturbations in Dynamics}
Let us start with the first design principle, namely robustness against
perturbations in dynamics. In other words,
we are aiming to design a system with a stable attractor.
For this end, we use Lyapunov exponent analysis and
a multicanonical Monte Carlo method~\cite{berg1991multicanonical,berg1992new}.

Once a network $W$ is given,
the finite time maximum Lyapunov
exponent $\lambda_{1}$~\cite{ott2002chaos} is calculated for the dynamical system in
Eq.~(\ref{eq:model}), starting from a given initial state
$x^{0*}_{i}$\footnote{Throughout this study, $x^{0*}_{i}=\sin (i)$ is used. We confirm that the
choice of the initial state does not affect the results.}.
We perform the simulation up to $T=1500$ and regard the first $T'=1000$
steps as transient and discard them.

Our aim here is to sample networks with negative $\lambda_{1}$, which indicates that dynamics of a network is stable.
Such networks are expected to be rare for large $N$, because dynamics tend to be chaotic at the present parameters.
We define the probability density of $\lambda_{1}$ as
\begin{equation*}
 D(\lambda)=\int_{0}^{1} \delta(\lambda _{1}(W)-\lambda)
  p(W) \ \ \Pi_{i, j}dW_{ij}, 
\end{equation*}
where $\delta$ is the Dirac $\delta$-function, and
$p(W)$ is the prior probability density that a network $W$ appears under random
sampling. We consider here a network ensemble in which $p(W)$ is a uniform distribution under the constraints of $W_{ii}=0$ and $\sum _j ^N W_{ij}=1$, given by 
\begin{eqnarray}
p(W)&\propto \Pi_{i}\delta(\sum_j W_{ij}-1)\delta(\sum_j
 \theta(W_{ij})-k)\delta(W_{ii}), \label{eq:measure}
\end{eqnarray}
where $\theta (z)$ is a function that satisfies $\theta (z)=1$ for $z\neq
0$ and $\theta (z)=0$ for $z=0$.

If a random sampling method is adopted in order to sample a network with ``rare'' value of $\lambda_{1}$ with probability $D(\lambda_{1})$, $1/D(\lambda_{1})$ samples are required at least.
If an annealing method or an steepest descent method is adopted instead, we would obtain only a network with negative 
$\lambda_{1}$ but could not estimate D($\lambda_{1}$), which plays a key role in the further analysis.
Alternatively, we apply multicanonical Monte
Carlo method~\cite{berg1991multicanonical,berg1992multicanonical,iba2001extended}, which has been used in fields of statistical physics, such as spin glass~\cite{wang2001efficient,wang2001determining} and other studies~\cite{saito2011probability,saito2010multicanonical}.
This method allows us to sample networks with
negative $\lambda_{1}$ efficiently and to estimate $D(\lambda_{1})$.

Our multicanonical Monte Carlo strategy adopted in this study is to perform
random walks in $\lambda_{1}$ space by generating a Markov chain, where each step is biased inversely proportional to the probability $D(\lambda_{1})$, and thereby it enables us to obtain a flat histogram in $\lambda_{1}$ space, namely 
to equally sample $\lambda_{1}$ whose $D(\lambda_{1})$ are many orders of magnitude different.
In order to generate the Markov chain in multicanonical Monte Carlo,
a key quantity is the weight function
$w(\lambda_{1})$ of $\lambda_{1}$. 
If we have $w(\lambda_{1})$ that is inversely proportional to $D(\lambda_{1})$, networks with various $\lambda_{1}$ value are generated one after another, using the Markov process described in the Appendix (i).
As a consequence, a uniform distribution of $\lambda_{1}$ (i.e., a flat histogram of $\lambda_{1}$) is obtained. 
We call these procedures as ``random walk in $\lambda_{1}$ space''. Details of the algorithm are given in the Appendix (i) and (ii).
However, neither $w(\lambda_{1})$ nor $D(\lambda_{1})$ are known a priori.
In this study, the Wang and Landau algorithm~\cite{wang2001efficient,wang2001determining} is
used to construct and to tune the weight function $w(\lambda_{1})$.
Details of the implementation are given in Appendix (ii).

Figure 1 shows the calculated densities
$D(\lambda_{1})$ of $\lambda_{1}$ for the fixed input degree $k=2-5$. 
Using density $D(\lambda_{1})$ in
Fig.~1, the probability
that networks with negative $\lambda_{1}$ are observed under random
sampling is calculated by
$P(\lambda_{1}<0)=\int _{\lambda_a} ^{0} D(\lambda_{1})d\lambda_{1}$.
We estimate $P(\lambda_{1}<0)$ with $k=2 - 5$ and
$k=N-1$, which are shown in Fig.~2.
Each $P(\lambda_{1}<0)$ shows that a stable attractor
becomes increasingly rare as $N$ or $k$ increases, indicating that these systems
are in the chaotic phase (we define that a system is in the chaotic phase when
only positive values of $\lambda_{1}$ appear as $N\to \infty$).
These results are consistent with the behavior of GCM with $(a,\epsilon)=(1.8,0.1)$~\cite{kaneko1990clustering}.

\section{Design of Robust Network Against Perturbations in Network}
Using the second design principle, we design systems that are robust against genotypic
perturbations (i.e., network perturbations). 
In other words, we design, using a multicanonical Monte Carlo method, networks $W$ whose
trajectory on the attractor hardly changes when a small network perturbation 
$\delta W$ is added.
We define the parameter sensitivity
and use it as a guiding function of the
robustness.

Sensitivity analysis using parameter sensitivity 
has been developed and applied in various fields~\cite{varma1999parametric,perumal2009dynamical}.
While most of these studies have dealt with continuous time systems,
we define the parameter sensitivity for discrete time systems as follows.

Let us denote the set of elements in $W$ in Eq.~(\ref{eq:model}) by a vector
$\ve{W}$. When a small network perturbation $\delta \ve{W}$ is
introduced into $\ve{W}$ at $t=T'$, the
displacement between unperturbed trajectory $\ve{x}^t(\ve{W})$ and
perturbed trajectory $\ve{x}^{t}(\ve{W}+\delta \ve{W})$ is approximated
by $(\partial \ve{x}^t/\partial \ve{W}) \delta \ve{W}$,
where $\partial \ve{x}^t /\partial \ve{W}$ is a $N \times N^2$
matrix. We call this matrix the sensitivity matrix $\Delta^t$, and
the time evolution of $\Delta^t$ is given by
\begin{equation*}
 \Delta^{t+1}=\frac{\partial \ve{F}}{\partial
  \ve{x}}\Delta^{t}+\frac{\partial \ve{F}}{\partial \ve{W}},
\label{eq:ps}
\end{equation*}
where $\partial \ve{F}/\partial \ve{x}$ is the Jacobian matrix and
$\partial \ve{F}/\partial \ve{W}$ is the parametric Jacobian
matrix. It should be noticed that the two trajectories $\ve{x}^t(\ve{W})$
and $\ve{x}^t(\ve{W}+\delta \ve{W})$ coincide for $t\le T'$, and thus $\Delta^{t}=0$ for $t\le T'$.
The growth rate of the displacement between $\ve{x}^t(\ve{W}+\delta \ve{W})$
and $\ve{x}^t(\ve{W})$ with respect to the perturbation vector 
$\delta \ve{W}$ is obtained by 
$\Delta ^t \ve{\delta \hat{W}} $, where $\ve{\delta \hat{W}}=\delta \ve{W}/|\delta \ve{W}|$.
The maximum value of $|\Delta ^t  \ve{\delta \hat{W}} |$ at time step $t$ is given by the maximum
singular value $\sigma_1 ^t$ of $\Delta ^t$ matrix.
$\sigma_1 ^t$ can be obtained by performing the singular value decomposition of
$\Delta ^t$.
$\sigma_1 ^t $ diverges for $t \to \infty$ when the maximum Lyapunov
exponent of the trajectory is positive. On the other hand, $\sigma_1 ^t$ oscillates or converges to
a constant value when the maximum Lyapunov exponent is negative.
Note that no parameters except for $W$ are
perturbed in this study.
Once a network $W$ is given, $\Delta ^t$ and its $\sigma _1^t$
are estimated for each time step.
We define parameter sensitivity $\gamma$ as the logarithm of an average of $\sigma_1 ^t$ along the trajectory:
 \begin{equation*}
	\gamma=\ln\left(\frac{\sum_{t=T'}^{T_{max}-1} \sigma_1
		    ^t}{T_{max}-T'}\right).
	\label{eq:ps2}
 \end{equation*}
Here, we regard the first $T'-1$ steps of the trajectory as transient,
and discard them. We use $T'=1000$ and $T_{max}=1500$.

Our goal is to sample networks with small $\gamma$.
However, networks with positive $\lambda_{1}$ tend to have large $\gamma$.
For this reason, and because almost all networks sampled by random sampling should have positive $\lambda_{1}$,
random sampling is not suitable for the sampling of networks with small $\gamma$.
Thus,  we again apply multicanonical Monte Carlo and perform random walks in $\lambda_{1}$ space with the same $w(\lambda_{1})$ estimated above.
These random walks facilitate efficient sampling of the networks with small $\gamma$, because networks with negative $\lambda_{1}$ are efficiently sampled.
We also obtain the two-dimensional density
$D(\lambda_{1}, \gamma)$ of $\lambda_{1}$ and $\gamma$ as follows:
we construct the two-dimensional histogram $h(\lambda_{1},\gamma)$
through the random walks, and, after $h(\lambda_{1},\gamma)$ is constructed, 
$D(\lambda_{1}, \gamma)$ is calculated by
       \begin{equation*}
	D(\lambda_{1},\gamma) \propto
	 \frac{h(\lambda_{1},\gamma)}{w(\lambda_{1})}.
	 \label{eq:reweight2d}
       \end{equation*}

Figure~3 shows $D(\lambda_{1},\gamma)$ for $N=10$ and $20$ with
$k=2$ and $5$.
These results show that although networks with small $\gamma$ can take various values for
$\lambda_1$, the vast majority of such networks with small $\gamma$ have
negative but near zero $\lambda_1$ (see red rectangles
in Fig.~3).
Figure~5 shows an example of an optimized network with small $\gamma$.
Although we have examined network topology of optimized networks sampled by multicanonical Monte Carlo,  
no characteristic difference was found in topology between networks with small $\gamma$ and those with large $\gamma$.

This indicates that, when robustness against network
perturbations is optimized (i.e., when $\gamma$ is minimized), networks with negative but near zero $\lambda_1$ will appear with high probability.
This appearance of systems with marginal stability can be
interpreted as self-organization of {\it the edge of
chaos}. In this scenario, the system automatically comes close to {\it the edge of
chaos} without tuning parameters.

In order to confirm that optimization of obustness against network perturbations leads
to the emergence of {\it the edge of
chaos}, we perform simulated annealing.
Here, we define a parameter sensitivity $\Gamma$ without the linear approximation:
\begin{equation*}
 \Gamma =\left\langle  \frac{\sum_{t=T'}^{T_{max}} |\ve{x}^t(\ve{W}+\ve{\delta
	   W}) - \ve{x}^t(\ve{W})|  }{T_{max}-T'} \right\rangle_{\delta \ve{W}},
\label{eq:Gamma}
\end{equation*}
where $\langle \rangle _{\delta \ve{W}}$ represents an average over
realization of $\delta \ve{W}$. We minimize
this $\Gamma$ using the simulated annealing.
The average is taken over 100 samples of 
$\delta \ve{W}$, and the parameters $T'=1000$ and $T_{max}=1500$ are used.
In each step of the simulated annealing, 
a transition from the current state $W$ to a proposed candidate $W'$ is accepted if and only if
the ratio $\exp[-\beta(\Gamma(W')-\Gamma(W))]$ is smaller than a random number uniformly distributed in $(0,1]$.
Here, temperature $1/\beta$ is lowered with the progress of simulation step $n$.
We choose this
       $\beta$ as $\beta=10 n/3$ (for $n<30000$) and $\beta=\infty$
       (for $n\ge30000$).
Note that the function $\Gamma $ that we aim to minimize
fluctuates due to the finite sample size of $\delta \ve{W}$, and thus
occasionally an inferior network could be accepted or a suitable
network could be rejected, even for $\beta=\infty$.

In Fig.~4, we plot $\lambda_{1}$ and $\Gamma$ for
networks that are sampled during the simulated annealing.
These results indicate that most networks obtained in the last half of
the simulations ($n \ge 30000$) are in the region $-0.05\le
\lambda_{1}<0$, and we
regard this region as {\it the edge of chaos}.

\section{Discussion}
In summary, using multicanonical Monte Carlo method, 
we have observed emergence of the systems at {\it the edge of chaos} as a
self-organization phenomenon 
with only the requirement of robustness against network perturbations, which can be interpreted as mutational robustness in the context of the gene regulatory network.
We have also performed simulated annealing and confirmed this scenario. 
We emphasize that no fine tuning of other parameters, such as number of input
links $k$ or model parameters $(a,\epsilon)$, is needed.
The emergence of the {\it edge of chaos} with the requirement of mutational
robustness is somehow counterintuitive because robustness against network perturbations (mutational robustness)
seems to be positively correlated with dynamical stability. 
The mechanism of the emergence of {\it the edge of chaos}, revealed by
a multicanonical Monte Carlo method, is as follows:
when mutational robustness is required, selected systems need to have
$\lambda_{1} \le 0$ because $\gamma$ for $\lambda_1 >0$ diverges as $t \to \infty$.
The density $D(\lambda _1)$ is an increasing function for $\lambda _1 \le 0$.
Therefore, the density of networks becomes largest at
$\lambda_{1}\sim 0$ under the condition $\lambda_{1}\le 0$ 
(see Fig.~1 and 3, red rectangles in Fig.~3 indicate the degeneracy of a large numbers of networks with small $\gamma$).
Due to this degeneracy, systems have a high probability of being at {\it
the edge of chaos}.

Similar results have been found in recent numerical
studies of gene regulatory network~\cite{bornholdt2000robustness,szejka2007evolution,sevim2008chaotic},
indicating that
systems that have the ability to reach a stable
fixed point with transient chaotic behavior appear with only the requirement of robustness against genotypic perturbations.
These results can be also interpreted as the emergence of {\it the edge of
chaos}. However, it has not until now been explained why such systems are selected.
In this paper, we have proposed a mechanism for the emergence of {\it the
edge of chaos}, namely that
the vast majority of networks that are robust against network perturbations have marginal
dynamical stability, and thus networks at the {\it edge of chaos} are selected when robustness against network perturbations is required.
Note that the converse is not necessarily true: networks with marginal dynamical stability are not always robust against network perturbations. 
It is reasonable to think that this degeneracy of marginally stable networks
appears whenever
parameters are set in the chaotic phase, in which chaotic systems are
obtained under random construction of systems for $N \to \infty$.
Based on the fact that similar results were found in the previous studies~\cite{bornholdt2000robustness,szejka2007evolution,sevim2008chaotic,torres2012criticality}, most of what we
discussed here seems not to depend on the details of the specific model.

Mutational robustness seems to be a natural requirement
for living systems.
The present study provides the following possible scenario for the emergence of {\it the edge of chaos}: the requirement of mutational
robustness drives organisms to {\it the edge of chaos}, whether or
not staying in such a regime is preferable for living systems.
In fact, several recent studies have suggested that gene networks of real organisms stay
at the {\it edge of chaos}~\cite{serra2007simple,shmulevich2005eukaryotic,nykter2008gene,balleza2008critical,chowdhury2010information}. 
We also expect that the scenario is also applicable for the explanation of criticality in neural dynamics~\cite{beggs2003neuronal,beggs2004neuronal,plenz2007organizing}, because synaptic connections seem to be designed so that neuron firing patterns do not change radically due to perturbations in the connection.  The concept presented here that robustness against network perturbations leads to the edge of chaos may also provide insights into the design of artificial robust networks.

\section*{Acknowledgements}
We would like to acknowledge encouragement and help from Kunihiko Keneko. 
This work was supported by the
Global COE Program (Core Research and Engineering of
Advanced Materials-Interdisciplinary Education Center for
Materials Science), MEXT, Japan. All simulations were performed
on a PC cluster at Cybermedia Center, Osaka University.

\newpage
\section*{References}
\bibliographystyle{unsrt}

\section*{Appendix: (i) details of the implementation of the Metropolis-Hastings algorithm}
We start from a randomly generated network $W$ with $N$ genes and $k$
input links for each gene. 
In each step of the Monte Carlo method, we generate a candidate of a new network $\ve{W}^{candidate}$ by
(1) exchanging two links or by (2) resampling interaction strengths.
We choose (1) or (2) with probability 1/2.
Process (1) (Exchanging links) is performed as follows:
a nonzero element of $W_{ij}$ is randomly chosen and the another
element in {\it j}-th row $W_{i'j}$ that satisfies both $i' \neq i$ and
$i' \neq j$ is chosen randomly and then these two links are exchanged. 
Process (2) (resampling interaction strengths) is performed as follows:
two nonzero elements $W_{ij}$ and $W_{i'j}$ are randomly chosen,
and then $W_{ij}$ and $W_{i'j}$ are replaced by resampled elements $W'_{ij}$
and $W'_{i'j}$, where $W'_{ij}$ is drawn from a random number uniformly distributed in
	    $(0,W_{ij}+W_{i'j}]$ and $W'_{i'j}$ is calculated by $W'_{i'j}=W_{ij}+W_{i'j}-W'_{ij}$.
After a new candidate $\ve{W}^{candidate}$ is obtained,
$\lambda_{1}(\ve{W}^{candidate})$ is calculated from the dynamical system
in Eq.~(1). A transition from the current state $\ve{W}^{current}$ to a proposed candidate $\ve{W}^{candidate}$ is accepted if and only if
the Metropolis ratio~\cite{metropolis1953equation,hastings1970monte}, $R=w(\lambda_{1}(\ve{W}^{candidate}))/w(\lambda_{1}(\ve{W}^{current}))$
is smaller than a random number uniformly distributed in $(0,1]$.
By this procedure, we update the current state to a new state $\ve{W^{new}}$.
Note that we adopt $\ve{W}^{current}$ as $\ve{W}^{new}$ when the $\ve{W}^{candidate}$ is rejected.

\section*{Appendix: (ii) details of the construction of $w(\lambda_{1})$ (the Wang-Landau algorithm)}
We divide the given prescribed interval $\lambda_a < \lambda_1< \lambda_b$ into small bins of width $\delta \lambda$, and 
assume that $w(\lambda_{1})$ is same within each bin.
We start from a uniform $w(\lambda_{1})$ in $\lambda_{a}<\lambda_{1}< \lambda_{b}$.
One step of the Metropolis - Hastings algorithm using $w(\lambda_{1})$ is performed as described in the Appendix (i), and then,  $w(\lambda_{1})$ is modified as $w(\lambda_{1}(\ve{W}^{new})) \to w(\lambda_{1}(\ve{W}^{new}))/ f$, where $f$ is the modification factor.
By this modification, $w(\lambda_{1})$ of frequently visited bins are made smaller, and thus the appearance of such $\lambda_{1}$ is suppressed.
The modification factor is set as $f=e$ at the beginning of the simulation, and the factor is gradually reduced through the simulation to approach to unity in the manner described in ~\cite{wang2001efficient,wang2001determining}.
We accumulate the number of samples in $\lambda_1 \sim \lambda_1+ \delta \lambda_1$ to make a histogram $h(\lambda_1)$.
The Metropolis - Hastings step and the modification of $w(\lambda_{1})$ are repeated until the histogram $h(\lambda_{1})$ is sufficiently flat in $\lambda_a < \lambda_1< \lambda_b$.
This procedure allows us to construct $w(\lambda_{1})$ that is inversely proportional to $D(\lambda_{1})$. 
Once such $w(\lambda_{1})$ is obtained, $w(\lambda_{1})$ is fixed. The Metropolis-Hastings algorithm using this $w(\lambda_{1})$ enables a uniform sampling in $\lambda_{1}$ space (i.e., a random walk in $\lambda_{1}$ space) by generating a Markov chain, because each step is biased proportional to $D(\lambda_{1})^{-1}$ and probability of the candidate $\lambda_1$ is proportional to $D(\lambda_{1})$.
Note that in this study we determine $\lambda_{a}$ and $\lambda_{b}$ so that $P(\lambda_{1}<0)$ is precisely calculated;
we adopt the criteria that $P(\lambda_{a})/P(\lambda_{1}=0) \sim 10^{-2}$ and  $P(\lambda_{b})/P(\lambda^{*}) \sim 10^{-2}$, where $\lambda^{*}$ is the value of $\lambda_{1}$ for the peak of $D(\lambda^{*})$ (see Fig.~1).

\section*{Figure Captions}
\subsection*{Figure 1:}
Densities of finite time Lyapunov exponent under
 random sampling of networks with input
 degree (a) $k=2$, (b) $k=3$, (c) $k=4$, (d) $k=5$. 
 (a) Densities are calculated in the prescribed interval
 $-0.28 < \lambda_{1} < 0.39$ for $N=4\sim 10$ and  $-0.1 < \lambda_{1}
 < 0.39$ for $N=20$ and $30$.   
 (b) Densities are calculated in the prescribed interval
 $-0.28 < \lambda_{1} < 0.35$ for $N=4\sim 14$ and  $-0.15 < \lambda_{1}
 < 0.35$ for $N=20$.
 (c) Densities are calculated in the prescribed interval
 $-0.28 < \lambda_{1} < 0.35$ for $N=6\sim 14$ and  $-0.15 < \lambda_{1}
 < 0.35$ for $N=20$.
 (d) Densities are calculated in the prescribed interval
 $-0.28 < \lambda_{1} < 0.35$ for $N=6$, $-0.28 < \lambda_{1} < 0.37$
 for $N=8\sim 10$, $-0.20 < \lambda_{1} < 0.37$ for $N=12\sim 14$ and  $-0.1 < \lambda_{1}
 < 0.37$ for $N=20$.\\
 \\
 \\
 \subsection*{Figure 2:}
 Network size dependence of the probability $P(\lambda_{1}< 0)$ of networks with
 negative $\lambda_{1}$.
The logarithms of $P(\lambda_{1}<0)$ for $k=2 - 5$ decrease linearly
 or slightly faster than linear as functions of $N$.
The logarithm of $P(\lambda_{1})$ for $k=N-1$ decreases quadratically.\\
 \\
 \\
 \subsection*{Figure 3:}
The two-dimensional density $D(\lambda_{1},\gamma)$ of
$\lambda_{1}$ and $\gamma$ for (a) $N=10$, $k=2$, (b) $N=10$, $k=5$ (c) $N=20$, $k=2$ (d) $N=20$, $k=5$. 
A large fraction of networks with small $\gamma$ degenerate in the region
 indicated by the red rectangles.
The black lines indicate $\lambda=0$.
It should be noted that $\gamma$ of $\lambda_{1}>0$ diverges, when $(T_{max}-T') \to \infty$.\\
 \\
 \\
 \subsection*{Figure 4:}
Time series of sensitivity $\Gamma$ (upper panel) and
 the maximum Lyapunov exponent $\lambda_{1}$ (lower panel) in the course of simulated annealing. 
The red and blue lines indicate results for
$N = 10$ with $k = 2$ and $N = 10$ with $k = 5$, respectively. 
$\lambda_1$ and $\Gamma$ are calculated in the same simulations. 
The inverse temperature is set to be $\beta=\infty$ in the last half of these simulations.\\
\\
\\
 \subsection*{Figure 5:}
An example of an optimized network with small $\gamma$ for $N=10$ and $k=2$.
$\lambda_{1}$ and $\gamma$ calculated from this network are $\lambda_{1}= -0.068$ 
and $\gamma = -0.008$, respectively.
Thickness of allows indicate strength of connection $W_{ij}$.\\
\\
\\
\newpage
\begin{figure}
\includegraphics{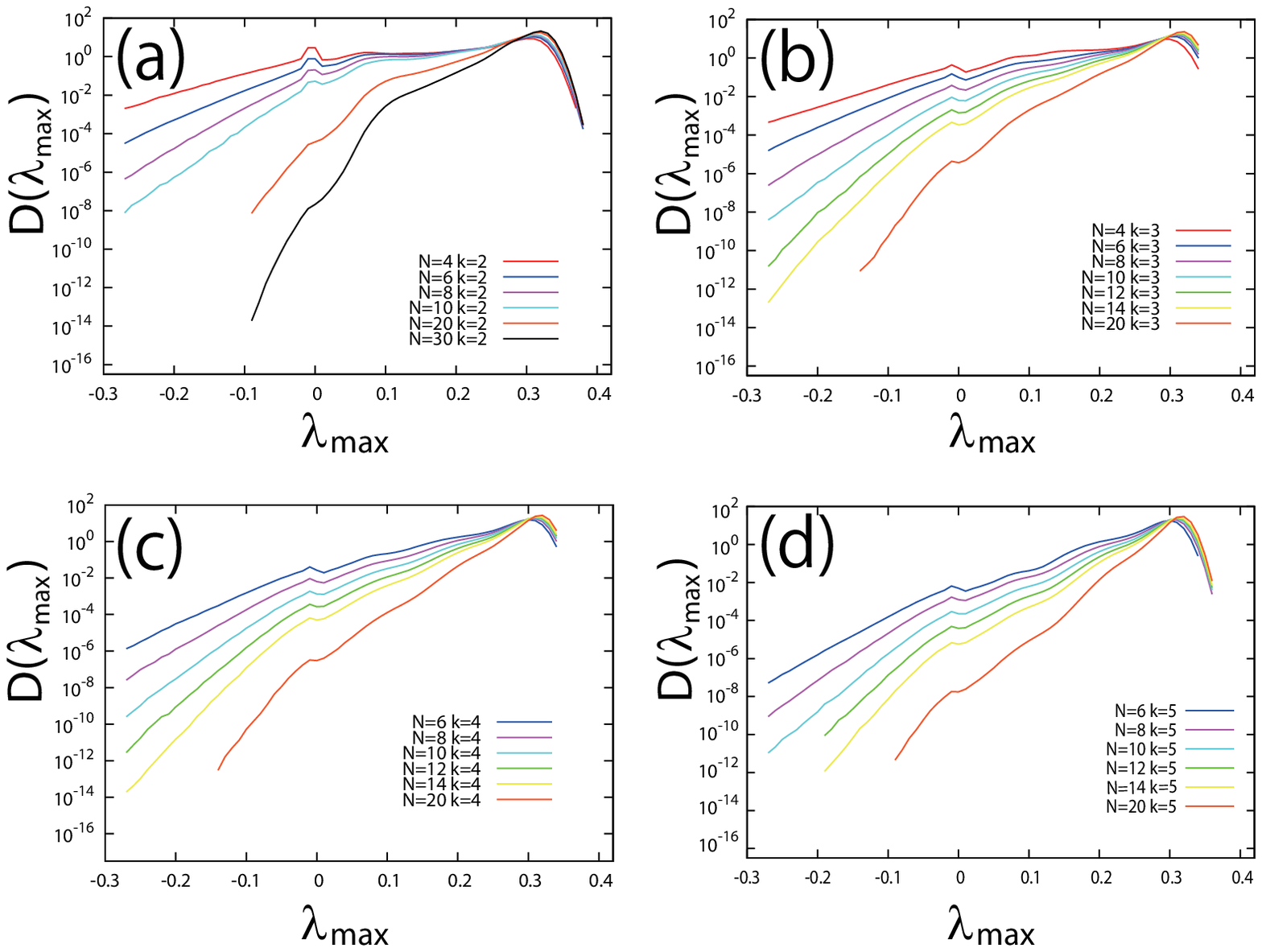}
\small
  \caption{  }
\end{figure}

\begin{figure}
\begin{center}
\includegraphics{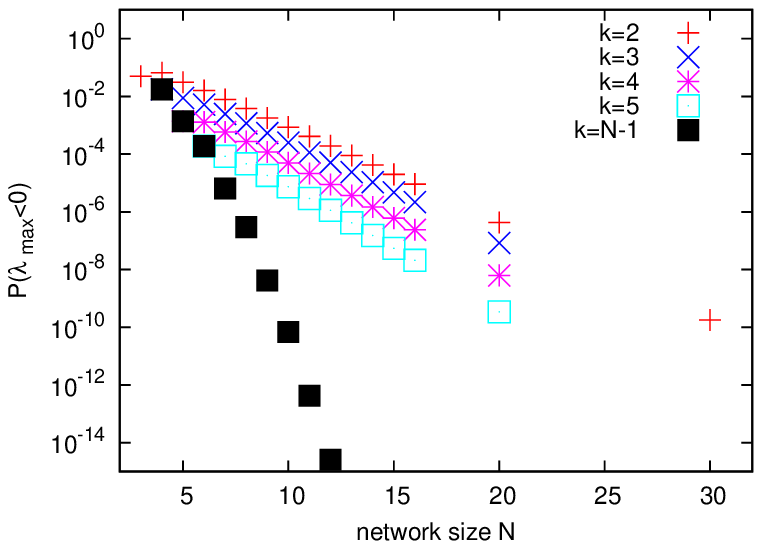}
\small
  \caption{ }
\end{center}  
\end{figure}

\begin{figure}
 \includegraphics{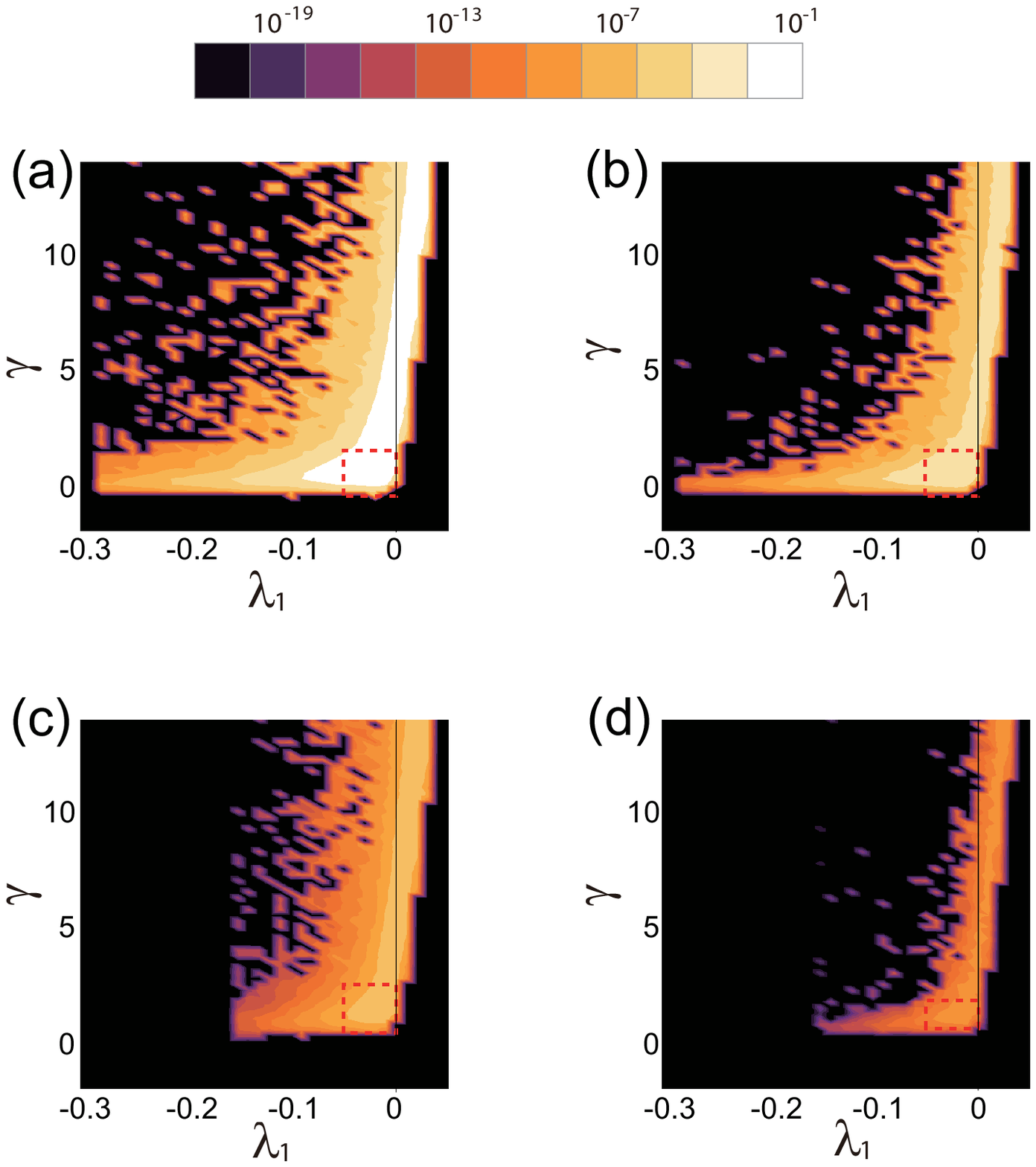}
\small
 \caption{ }
\end{figure}

\begin{figure}
\includegraphics{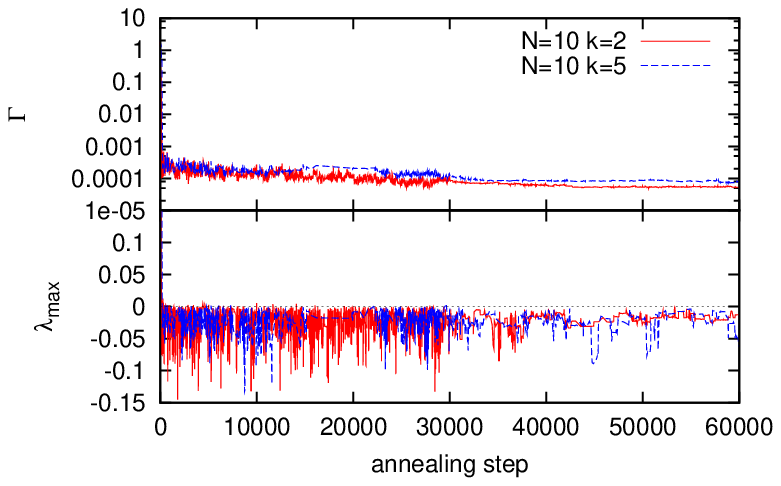}
\small
\caption{ }  
\end{figure}

\begin{figure}
\includegraphics{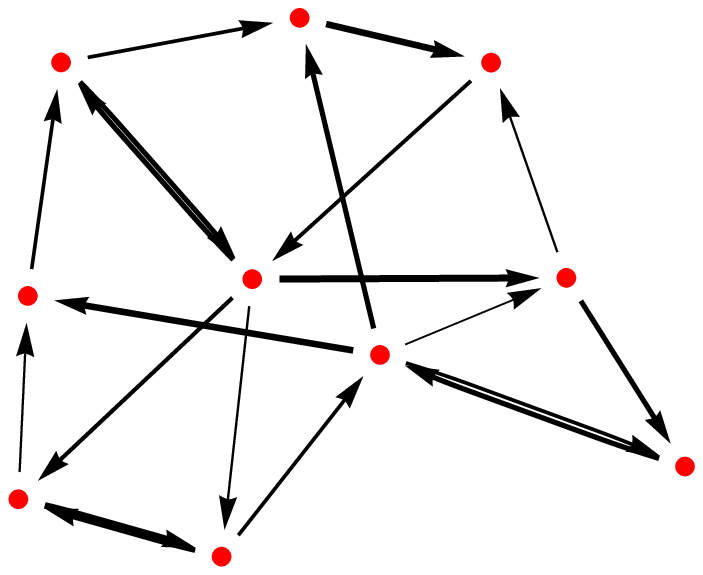}
\small
\caption{ }  
\end{figure}

\end{document}